\documentclass[
  journal=pasa,
  manuscript=research-paper, 
  year=2020,
  volume=37,
]{cup-journal}

\usepackage{microtype,siunitx,booktabs}
\usepackage{amsmath}	
\usepackage{amssymb}	
\usepackage{xcolor}
\newcommand\ed[1]{{\color{black}{#1}}}

\title{Beamforming approaches toward detecting the 21-cm global signal from Cosmic Dawn with radio array telescopes}

\author{D. C. Price}
\affiliation{International Centre for Radio Astronomy Research, Curtin University, Bentley, WA 6102, Australia}
\email[D. C. Price]{danny.price@curtin.edu.au}

\doi{10.1017/pasa.2020.32}

\received {dd Mmm YYYY}
\revised  {dd Mmm YYYY}
\accepted {dd Mmm YYYY}
\published{22 September 2020}


\newcommand{\tblnote}[1]{\multicolumn{2}{l}{\footnotesize{#1}}}

\begin{document}

\begin{abstract}
The formation of the first stars and galaxies during `Cosmic Dawn' is thought to have imparted a faint signal onto the 21-cm spin temperature from atomic Hydrogen gas in the early Universe. Observationally, an absorption feature should be measurable as a frequency-dependence in the sky-averaged (i.e. global) temperature at meter wavelengths. This signal should be separable from the smooth---but orders of magnitude brighter---foregrounds by jointly fitting a log-polynomial and absorption trough to radiometer spectra. A majority of approaches to measure the global 21-cm signal use radiometer systems on dipole-like antennas. Here, we argue that beamforming-based methods may allow radio arrays to measure the global 21-cm signal. We simulate an end-to-end drift-scan observation of the radio sky at 50--100\,MHz using a zenith-phased array, and find that the complex sidelobe structure introduces a significant frequency-dependent systematic. However, the $\lambda/D$ evolution of the beam width with frequency does not confound detection. We conclude that a beamformed array with a median sidelobe level $\sim-50$\,dB may offer an alternative method to measure the global 21-cm signal. This level is achievable by arrays with $O(10^5)$ antennas. 
\end{abstract}

\section{INTRODUCTION }
\label{sec:int}

Cosmic Dawn is a cosmological epoch spanning from the formation of the first stars ($z\sim30$) until reionization started in the intergalactic medium ($z\sim10$). The physics and cosmology of this epoch is imprinted upon the spin temperature of neutral Hydrogen gas (HI), making the detection of HI 21-cm emission from the Cosmic Dawn one of the best probes of the early Universe \citep{Furlanetto:2006,Morales:2010,Pritchard:2012,Cohen:2017}. 

Analogous to the cosmic microwave background (CMB), the 21-cm emission from Cosmic Dawn is expected to be measurable as a global, sky-averaged signal: an absorption trough in the global sky spectrum at observing frequencies of 45--130\,MHz, corresponding to redshifts pf $30 \gtrsim z \gtrsim 10$ \citep{Shaver:1999,Furlanetto:2006,Morales:2010,Pritchard:2012}. The depth, width, and central frequency of the absorption trough depend on astrophysical processes and cosmology, but it is expected to have a magnitude of $O(100)$\,mK and $O(10)$\,MHz width \citep{Cohen:2017}. This $\sim$100\,mK signal must be separated from astrophysical foregrounds, which are dominated by synchrotron emission that is many orders of magnitude brighter. Fortunately, synchotron emission is spectrally smooth, and can be accounted for using a low-order polynomial in log-frequency space. The 21-cm signal can therefore be recovered by jointly fitting a low-order polynomial and Gaussian absorption feature to measured data. 

Using this approach, the Experiment to Detect the Global EoR Signal (EDGES) reported the detection of a $\sim$18.7 MHz wide, $\sim$530 mK absorption feature at 78.1\,MHz \citep{Bowman:2018}. Remarkably, this signal is 2--3 times brighter than that expected from optimistic models \citep{Cohen:2017}, which would point toward gas temperatures during Cosmic Dawn being far cooler than predicted. However, there is ongoing debate as to whether or not the EDGES detection is bona fide, or the result of unmodelled systematics. \citet{Hills:2018} and \citet{Singh:2019} suggest that the feature may be explained by an unaccounted for sinusoidal ripple within the EDGES receiver; \citet{Bradley:2019} suggests that the finite ground plane under the antenna could give rise to a resonant spectral structure. \citet{Sims:2020} show via a Bayesian reanalysis that models including previously unmodelled systematics are decisively preferred, but do not rule out models which include a 21-cm signal with a revised $\sim$200\,mK amplitude (in closer agreement with predictions). Most concerning is that results from the Shaped Antenna Measurement of the Background Radio Spectrum 3 (SARAS 3) experiment reject the EDGES best-fit profile to 95.3\% \citep{Singh:2022}.

\begin{table*}
    \begin{center}
    \small
    \begin{tabular}{ l l }
    \hline
    Acronym & Experiment \\
    \hline
    EDGES-hi & Experiment to Detect the Global EoR Signal$^1$ \\
    EDGES-lo & Experiment to Detect the Global EoR Signal, low-band$^2$  \\
    BIGHORNS & Broadband Instrument for Global HydrOgen ReioNisation Signal$^3$ \\
    LEDA & Large-aperture Experiment to detect the Dark Ages$^4$  \\
    SARAS & Shaped Antenna measurement of the background RAdio Spectrum$^5$  \\
    SCI-HI & Sonda Cosmológica de las Islas para la Detección de Hidrógeno Neutro$^6$ \\
    PRIZM &  Probing Radio Intensity at high-$z$ from Marion$^7$\\
    REACH & Radio Experiment for the Analysis of Cosmic Hydrogen$^8$ \\
    DARE & Dark Ages Radio Explorer$^9$ \\
    \hline
    \tblnote{$^1$\cite{Rogers:2008, Bowman:2010, Rogers:2012}} \\
    \tblnote{$^2$\cite{Bowman:2018} $^3$\cite{Sokolowski:2015} $^4$\cite{Greenhill:2012, Price:2018} } \\
    \tblnote{$^5$\cite{Patra:2013, Singh:2018, Singh:2022} $^6$\cite{Voytek:2014} $^7$\cite{Philip:2019}} \\
    \tblnote{$^8$\cite{Cumner:2022} $^9$\cite{Burns:2021}} \\
    
    \end{tabular}
    \end{center}
    \raggedright
    
    \caption{Radiometer experiments to detect the global 21-cm signal.}
    \end{table*}

This tension, and difficulty in ruling out low-level systematic effects, highlights the need for alternative experimental approaches. A majority of global 21-cm  experiments, including EDGES, use a simple dipole-style antenna connected to an exquisitely calibrated receiver (see Table 1). One alternative approach is to use radio interferometers. Under normal assumptions and conditions, interferometers are insensitive to the global signal, as they do not sample the origin of the $uv$-plane. However, the interferometric response of a pair of closely-spaced antennas to the global signal is not zero \citep{Vedantham:2015,Presley:2015,Singh:2015}, and efforts to exploit this to detect the 21-cm signal are underway \citep{McKinley:2020,Thekkeppattu:2022}. For these experiments, effects such as mutual coupling and cross-talk must be understood and accounted for. An alternative method, in which the occultation of the Moon in front of a patch of sky is used to make a differential measurement of the 21-cm signal \citep{Vedantham:2015, McKinley:2018}. This approach suffers from comparably low sensitivity and is susceptible to reflected radio interference from Earth, although improved techniques and instruments may solve these.

Another observing strategy for radio arrays is to use beamforming techniques. In \citet{Dilullo:2020} and \citet{Dilullo:2021}, the Long Wavelength Array at Sevilleta (LWA-SV) was used in beamforming mode to simultaneously observe a calibrator source (Virgo A) and a cold patch of sky. The main advantages of this multi-beam technique is are a) a calibrator can be observed at the same time as a science target field to counteract time-dependent systematics, b) a cold patch of sky can be chosen, decreasing overall radiometer noise, and c) the foreground model within the beam is much simpler, by virtue of the smaller angular size. While this approach shows promise, systematics such as the ionosphere, and uncertainty in the beam response, are challenges that must be overcome.

Indeed, the complex beam response of a radio array is one of the primary arguments put forth against their use. Any variations of an antenna's beam power pattern as a function of frequency will lead to chromatic mixing of spatial structure into spectral structure \citep{Vedantham:2014, Bernardi:2015, Mozdzen:2017}. If the sky's spatial structure is not taken into account, or if the beam pattern is not known to enough accuracy, an unwanted spectral structure will be imprinted upon calibrated spectrum. Given an array of diameter $D$, the width of its primary beam is proportional to observing wavelength by $\lambda/D$. Consequently, there is considerable evolution of the beam across 45--130 MHz. Because of this, \citet{Dilullo:2021} employ a novel achromatic beamforming technique to counteract the evolution of the beam with increased frequency.

The frequency-dependence of an antenna beam is often referred to as beam chromaticity. The mixing of spatial structures into the observed spectrum requires a more complex treatment of foregrounds. \citet{Bernardi:2015} showed that beam chromaticity for a realistic dipole introduces spectral structure, but that it could be accounted for by using a higher-order polynomial fit. This concern drove EDGES to change their antenna to a simpler design that could be more accurately modelled in electromagnetic simulation packages \citep{Mozdzen:2016, Monsalve:2017a}. For the same reason, SARAS introduced new antennas at each iteration \citep{Patra:2013, Singh:2018, Singh:2022}.  High-gain antennas were considered for the upcoming REACH experiment, but simulations showed that their complex beam patterns resulted in larger residuals than low-gain antenna designs \citep{Cumner:2022}.

As presented by \citet{Hibbard:2020}, beam chromaticity can be accounted for using eigenvectors formed from beam-weighted foregrounds. In agreement with previous findings \citep{Vedantham:2014, Bernardi:2015, Monsalve:2017}, the eigenvectors are shown to be more dependent upon the beam model than the foreground model. The methodology offers a practical approach to generating accurate beam-weighted foreground models that would be required for beamforming approaches.

In this paper, we use simulations to investigate how sidelobes and $\lambda/D$ frequency dependence of a beamformed beam affect global 21-cm signal experiments. For a $\sim$degree-sized beam, we find that frequency dependence of the primary beam does not confound detection; rather, the sidelobe structure introduces complex spectral features. If beam sidelobes can be reduced to $\sim$50 dB of the primary beam, a detection of the global 21-cm signal may be possible. 


\section{METHODS}
\label{sec:obs}

\begin{figure}[t]
\centering
\includegraphics[width=\textwidth]{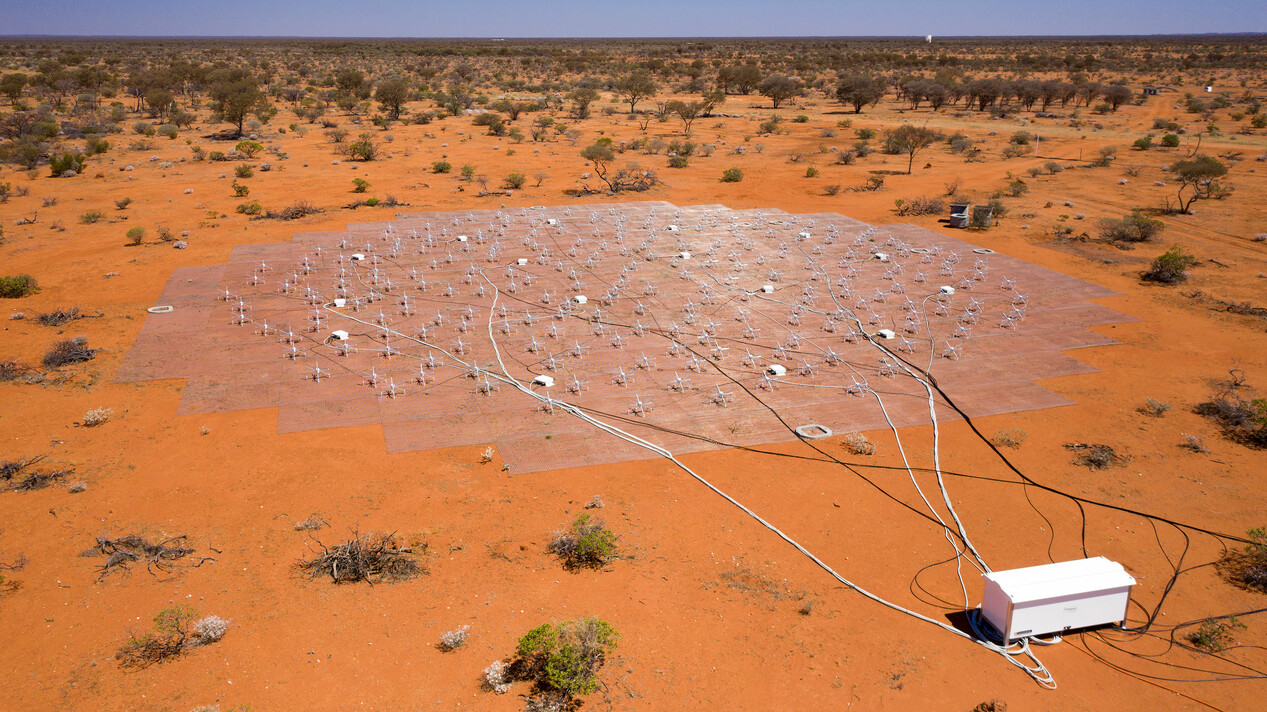}
\caption{An aerial view of the Engineering Development Array v2 (EDA2).}
\label{fig:eda}
\end{figure}

We investigate the plausibility of beamformer-based methods, by simulating observations that incorporated realistic beam patterns and sky foregrounds. Here, we simulated 24-hour drift-scan observations across 50--100\,MHz. While beamformed arrays allow for alternative pointing strategies, a drift-scan with the array pointed toward zenith is the most similar to current radiometer approaches. \ed{Further, by fixing to zenith, projection effects and effects due inter-antenna beam pattern variations are minimized.}

\begin{figure}[t]
\centering
\includegraphics[width=\textwidth]{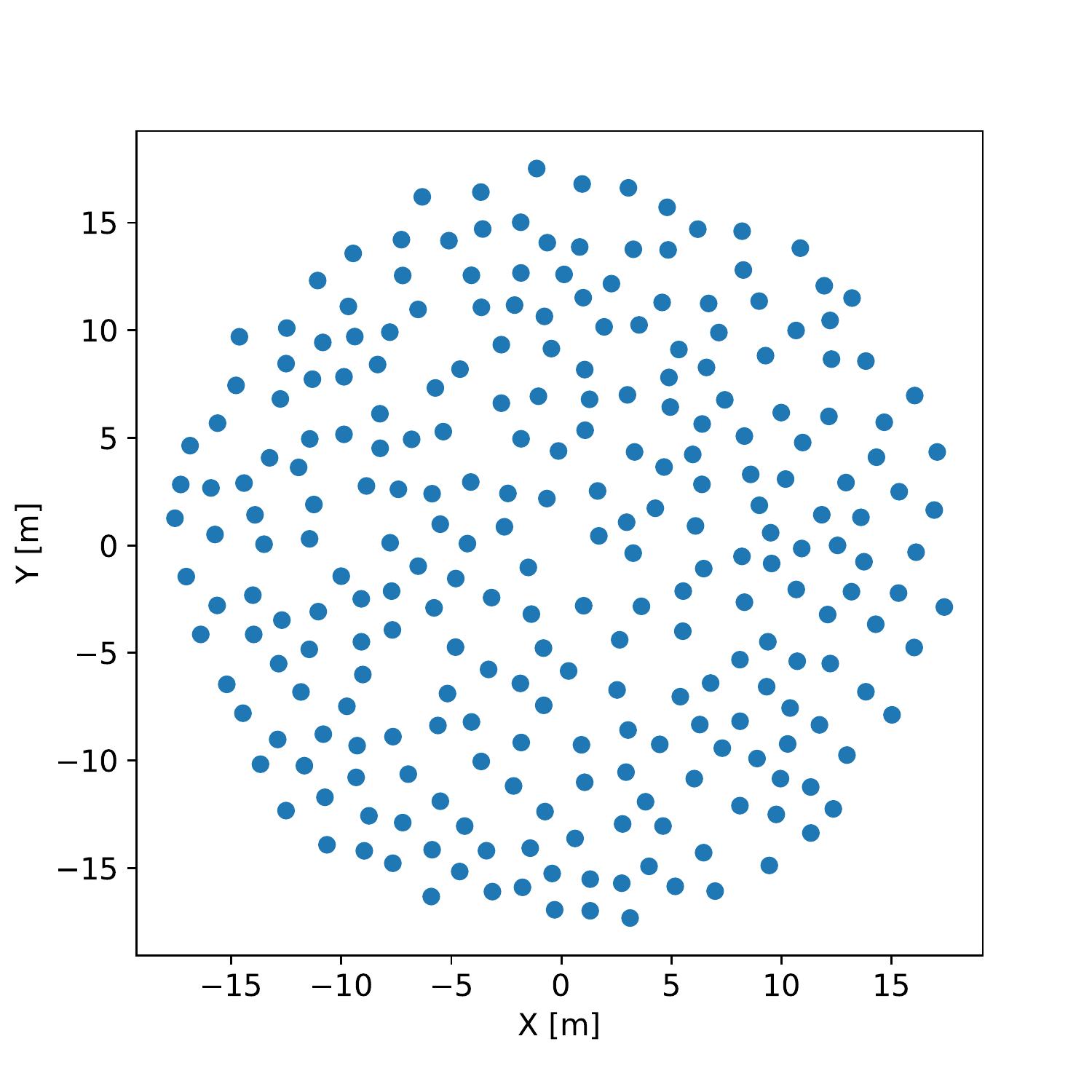}
\caption{EDA2 antenna layout, as used in OSKAR simulations.}
\label{fig:eda-layout}
\end{figure}

Our aim is to model the antenna temperature of a beamformed array. This is given by the average of the sky brightness $T_{\rm{sky}}(\theta,\phi,\nu)$
as seen from the antenna's location, weighted by the antenna's gain pattern
$B(\theta,\phi,\nu)$:
\begin{equation}
T_{\rm{ant}}(\nu)=\frac{\int d\Omega\, B(\theta,\phi,\nu)T_{\rm{sky}}(\theta,\phi,\nu)}{\int d\Omega\, B(\theta,\phi,\nu)}.\label{eq:sky}
\end{equation}
Separating the $T_{\rm{sky}}$ into a `foreground' component, $T_{\rm{fg}}$, and a cosmological term $T_{\rm{cosmo}}$ consisting of the sky-averaged 21-cm emission and the CMB background, we have
\begin{equation}
T_{\rm{ant}}(\nu)=\frac{\int d\Omega\, B(\theta,\phi,\nu)T_{\rm{fg}}(\theta,\phi,\nu)}{\int d\Omega\, B(\theta,\phi,\nu)} + T_{\rm{cosmo}}(\nu).\label{eq:fgbg}
\end{equation}
We chose to use the antenna layout of Engineering Development Array \citep[EDA,][]{Wayth:2017, Wayth:2021} as a primary example (Figs.\,\ref{fig:eda} and \ref{fig:eda-layout}). The EDA is a precursor instrument for the Square Kilometre Array SKA-low aperture array, which will operate across 50--350\,MHz. We chose the EDA2 as our example as is under active development, has a well-understood beam response, and is located close to EDGES within the Murchison Radio Observatory.

\subsection{Station beam}

We used the OSKAR\footnote{\url{https://github.com/OxfordSKA/OSKAR}} aperture array simulator \citep{Dulwich:2009} to simulate the beam pattern of EDA2. OSKAR was designed to produce simulated data from aperture array telescopes that employ beamforming, interferometry, or both methods. OSKAR employs the Radio Interferometer Measurement Equation \citep[RIME,][]{Hamaker:1996,Smirnov:2011} to produce simulated visibility data. In the RIME framework, beamforming---the weighted sum of the response from each antenna with complex beamforming weight---is equivalent to computing a `station beam' Jones matrix.

\begin{figure}
\centering
\includegraphics[width=\textwidth]{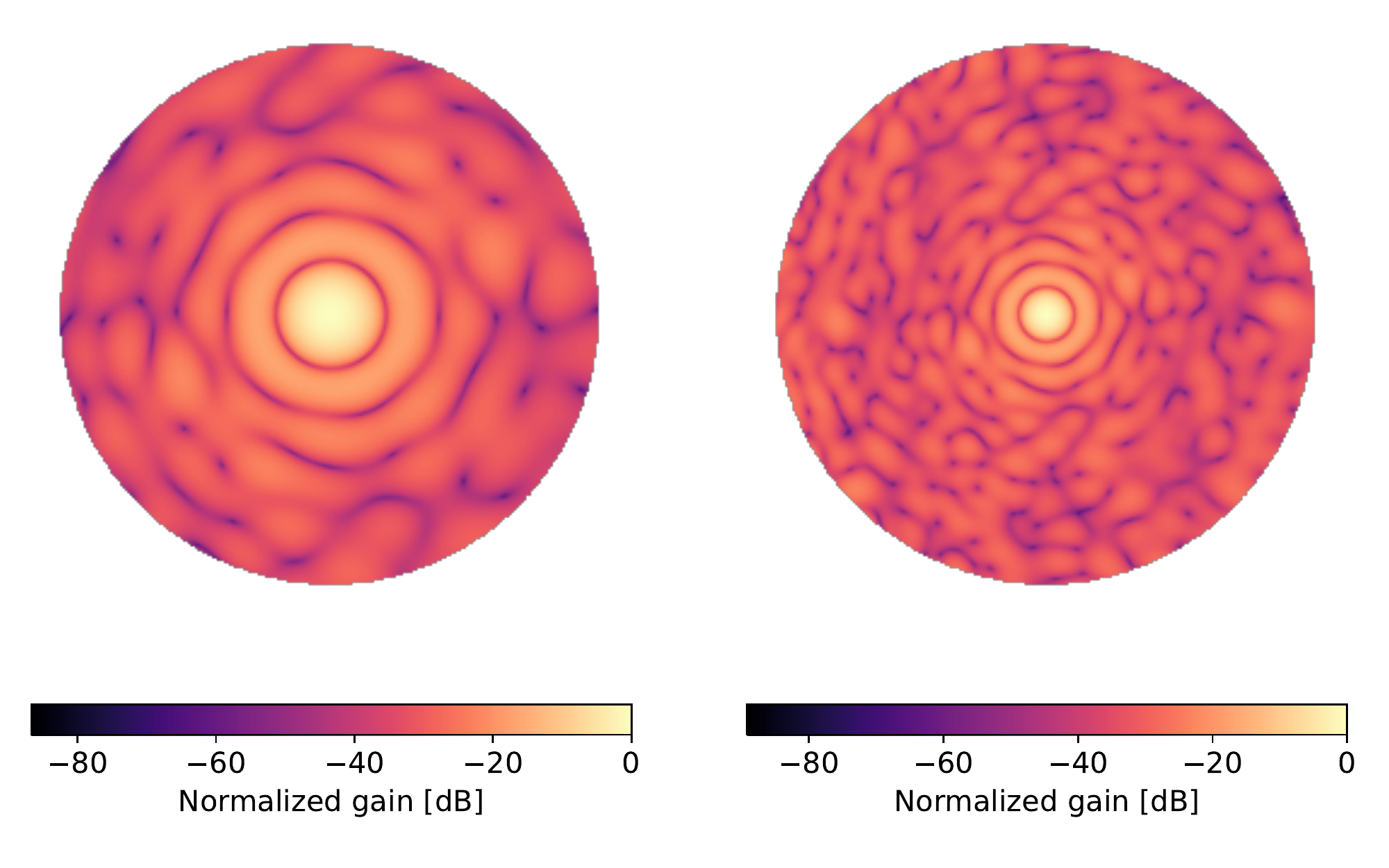}
\caption{Simulated station beam patterns for EDA2 (all-sky orthographic projection), at 50\,MHz (left) and 100\,MHz (right).}
\label{fig:ant-patterns}
\end{figure}

\begin{figure}
\centering
\includegraphics[width=\textwidth]{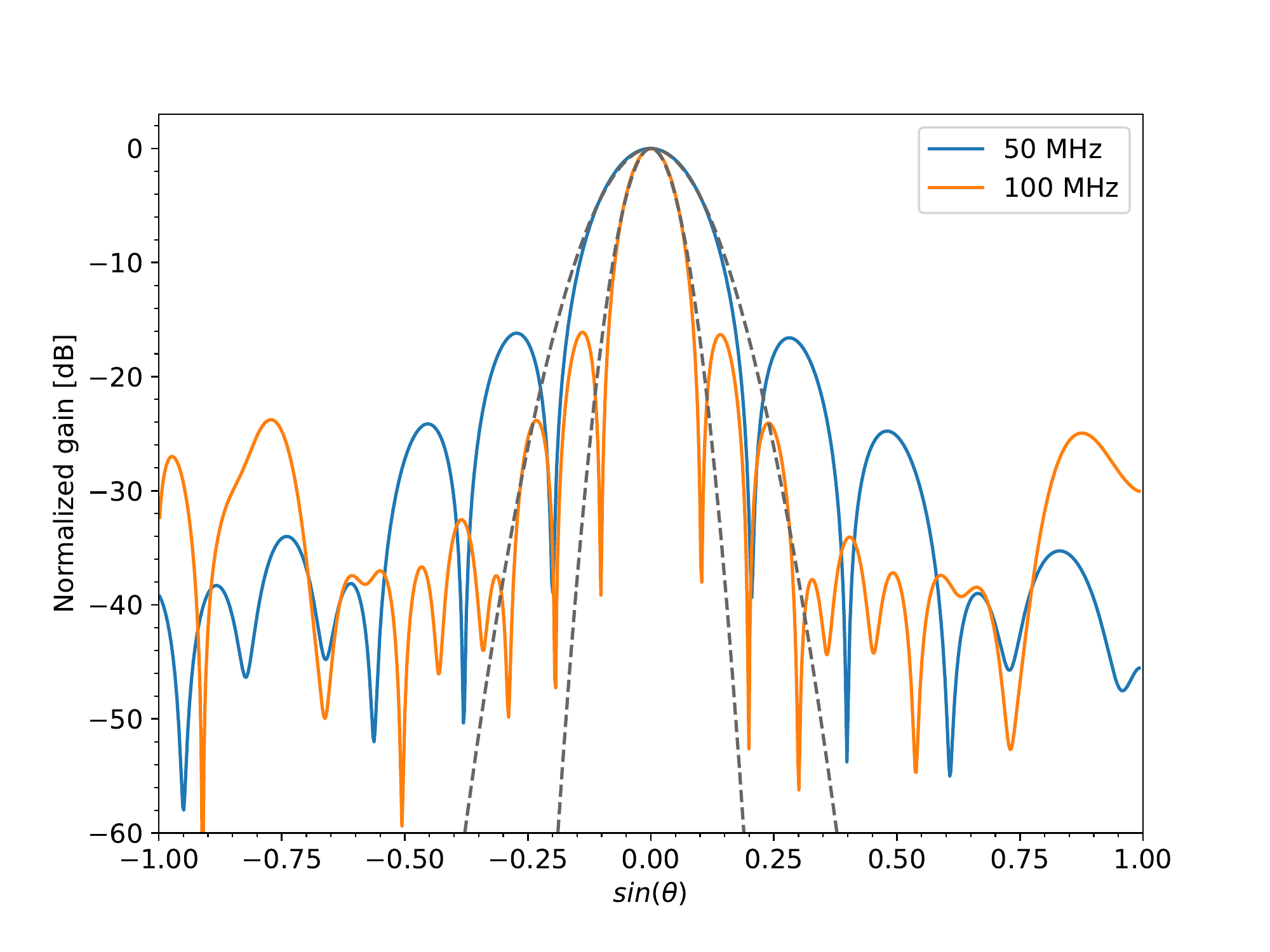}
\caption{Cuts of antenna gain patterns for EDA2, at 50 and 100\,MHz. Gaussian fits to the primary beam are shown as dashed grey lines.}
\label{fig:ant-patterns-2d}
\end{figure}

We used the OSKAR task {\texttt{oskar\_sim\_beam\_pattern}} to generate zenith-phased station beams across 50--100\,MHz in 1\,MHz steps, and saved these in FITS format \citep{Wells:1981}. Simulated station beams at 50\,MHz and 100\,MHz are shown in Fig.\,\ref{fig:ant-patterns}. Evident in Fig.\,\ref{fig:ant-patterns} is both the scaling of beam width with wavelength $\lambda/D$, and the complex sidelobe response. Cuts through the beam centres are shown in Fig.\,\ref{fig:ant-patterns-2d}, with Gaussian fits to the primary beams overlaid. The first sidelobe peaks at $\sim15$\,dB below the primary beam response.

\subsection{Astrophysical Foregrounds}

\begin{figure*}
\centering
\includegraphics[width=12cm]{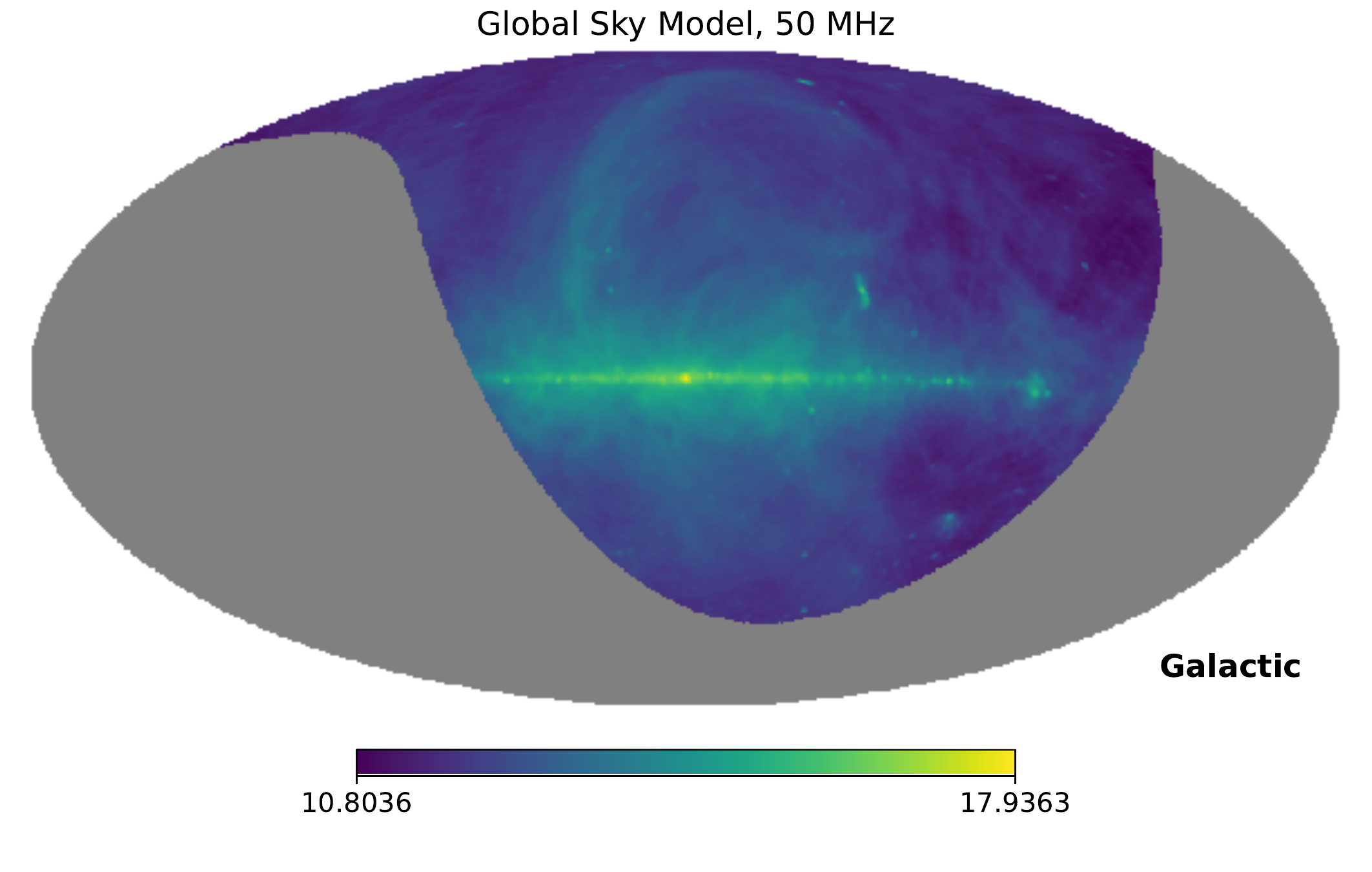}
\caption{Diffuse sky model at 50 MHz, generated by PyGDSM using the \citet{deOliveira-Costa:2008} GSM. Colorscale is $log_{2}(T)$. The sky area not visible to EDA2 has been masked.}
\label{fig:gsm}
\end{figure*}

\ed{In order to} simulate astrophysical foregrounds, we used the \textsc{PyGDSM} package \citep{Price:2016}, which provides an interface to three global diffuse sky models (GDSMs): 
\begin{itemize}
    \item \textit{GSM08}: The Global Sky Model \citep{deOliveira-Costa:2008}.
    \item \textit{GSM17}: The `improved' Global Sky Model \citep{Zheng:2017}.
    \item \textit{LFSM}: The Low Frequency Sky Model \citep{Dowell:2017}.
\end{itemize}
All three GDSMs can be used to estimate the diffuse foregrounds across the 50--100\,MHz range. 

During our investigations we noted that spectra generated with GSM17 showed a discontinuity around 45\,MHz, which we believe arises due to a scale offset in the underlying \citet{Guzman:2011} map at 45\,MHz. Indeed, \citet{Monsalve:2021} reports a $-160\pm78$\,K zero-level correction is required for the \citet{Guzman:2011} map, based on EDGES measurements of the sky spectra. A similar correction was applied to the LFSM to remove an inflection at 45\,MHz  (J. Dowell, personal comms). Regardless, the models show broad agreement to within $\sim10\%$. For comparison with \citet{Bernardi:2015}, we used the GSM08, and generated sky brightness maps (Fig.\,\ref{fig:gsm}) over  50--100\,MHz in 1\,MHz in HEALpix format \citep{Gorski:2005} with {\texttt{NSIDE=256}}.

\subsection{Drift scan simulation}

\ed{The antenna temperature} for a given pointing was simulated by combining HEALpix maps generated with \textsc{PyGDSM} for $T_{\rm{fg}}(\theta, \phi, \nu)$, and beam models from \textsc{OSKAR} for $B(\theta, \phi, \nu)$. \ed{We then} investigated the effect of the sidelobes by comparison against a simple Gaussian beam fitted to the primary beam (Fig.\,\ref{fig:ant-patterns-2d}). 

The zenith-pointed OSKAR beam models are output in equatorial coordinates. To simulate a drift scan, we converted the beam patterns celestial coordinates corresponding to zenith at 48 equally-spaced times across a sidereal day, then regridded into HEALpix arrays with \texttt{NSIDE=256}. $T_{\rm{cosmo}}$, the 21-cm absorption trough, is modelled as a Gaussian with 5\,MHz and 200\,mK amplitude, centered at 78\,MHz. These values are based on EDGES, using the lower revised amplitude from \citet{Sims:2020}. We then numerically computed $T_{\rm{ant}}(\nu)$ (Eq.\,\ref{eq:fgbg}).

Fig.\,\ref{fig:driftscan} shows the resulting simulated drift-scan observations for the EDA2. In comparison to a dipole, the directional beam of EDA2 results in lower $T_{\rm{ant}}$ for most LSTs due to the suppression of the hotter Galactic plane foregrounds.

\section{RESULTS}
\label{sec:results}

\begin{figure*}
\centering
\includegraphics[width=16cm]{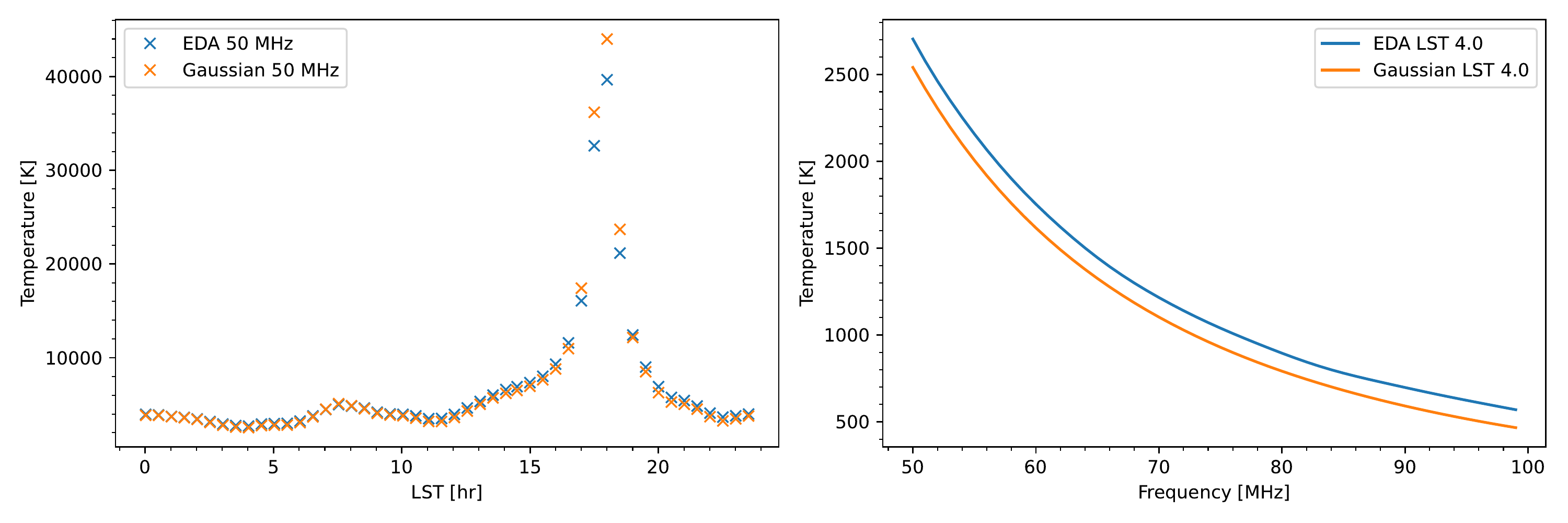}
\caption{Simulated observations with the EDA2, using a simulated beam model (blue) and Gaussian beam model (orange).}
\label{fig:driftscan}
\end{figure*}

\begin{figure}
\centering
\includegraphics[width=\textwidth]{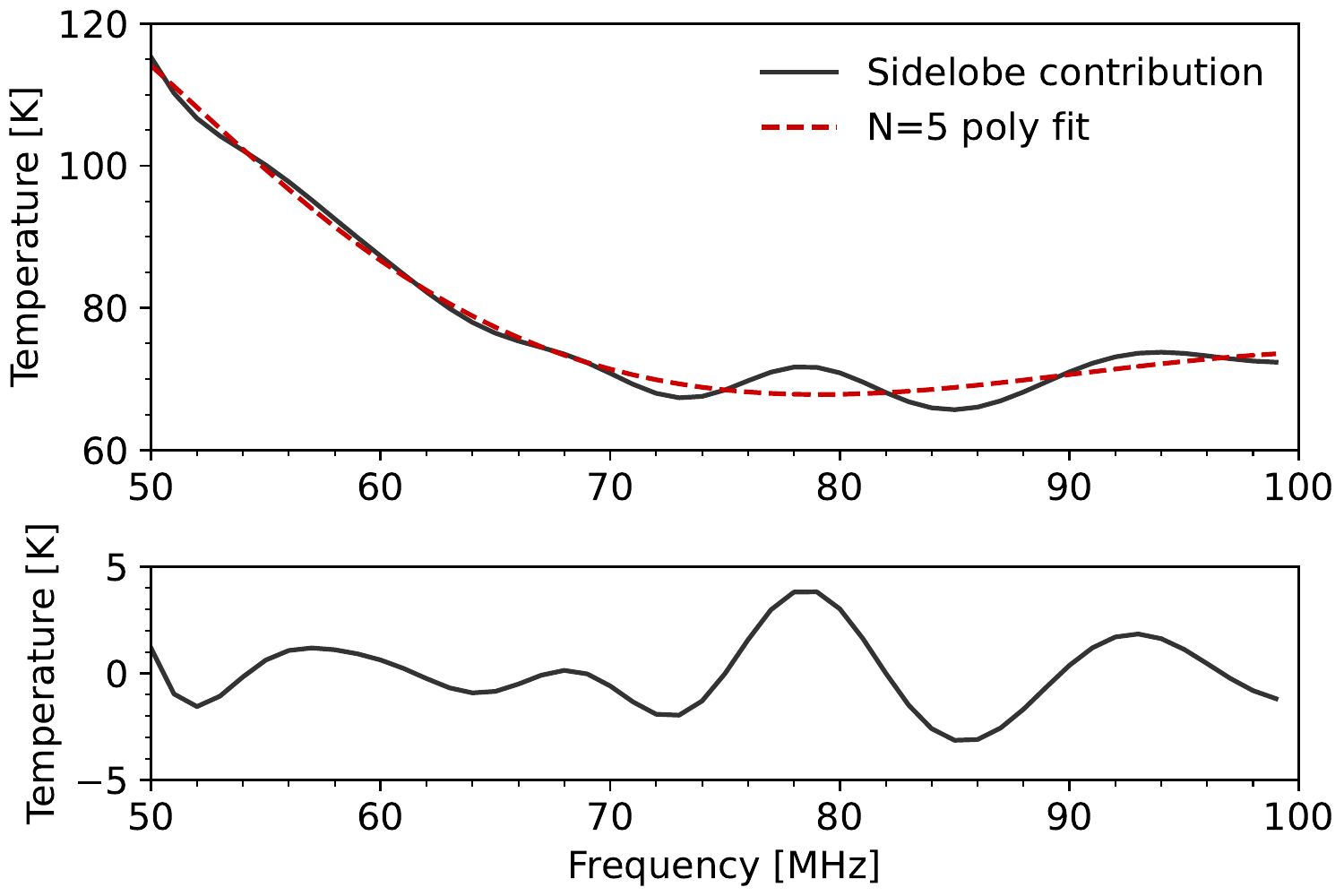}
\caption{Contribution of sidelobes to antenna temperature, \ed{as given by the difference between simulated and Gaussian beam models shown in Fig 6}, at LST=4 hr (top panel, black curve), with $N$=5 log-polynomial fit (dashed red). The residual after subtraction of the fit is shown in the bottom panel. }
\label{fig:driftscan-diff}
\end{figure}

Our drift-scan simulation of EDA2 (Fig.\,\ref{fig:driftscan}) shows that antenna temperature changes markedly with sidereal time, from a minimum of 2703\,K at 50 MHz (4 hr LST) to a maximum of 39,650\,K as the Galactic plane crosses the primary beam (LSTs 15--20 hr). In contrast, EDGES measures a much more gradual change in $T_{\rm{ant}}$ with LST 
\citep[see Tab.2,][]{Monsalve:2021}, reaching a maximum 16,375$\pm$308\,K at 45\,MHz at LST 17.84 hr, and minimum 5607 $\pm$ 107\,K at LST 2.56 hr. The lower antenna temperature of the beamformer approach would mean that a target radiometer noise level can be reached within a shorter integration time. 

Nevertheless, the sidelobe response of EDA2 introduces a chromatic systematic that depends upon both frequency and time. In the right panel of Fig.\,\ref{fig:driftscan}, the antenna temperature $T_{\rm{ant}}(\nu)$ at an LST of 4.0\,hr is shown, for both EDA2 and an ideal Gaussian beam with the same width. The contribution of the antenna temperature due to the sidelobes is given by the difference between the two spectra (Fig.\,\ref{fig:driftscan-diff}). The sidelobes introduce complex frequency structure, which remain after a $N$=5 log-polynomial fit. The systematic term shown in Fig.\,\ref{fig:driftscan-diff} has a 1.6\,K RMS: larger than the reported 500\,mK EDGES 21-cm signal. 

We further investigated the effect of the sidelobes, by modifying the EDA2 beam pattern to suppress the sidelobes by 10\,dB and 20\,dB (i.e. factors of 10 and 100) and re-running the simulation. \ed{The sidelobes were first isolated by fitting and subtracting the primary beam from the beam pattern. We then multipled the sidelobes by an attenuation factor, then added the result back to the primary beam.} The residuals for a $N=5$ log-polynomial fit are reduced to 208\,mK RMS and 20\,mK RMS, respectively. The markedly lower residuals suggest that sidelobes are a far larger source of systematic noise than the  $\lambda/D$ evolution of beam width. 

To see if a the 21-cm absorption feature can be extracted from the simulated antenna temperature, we used \textsc{HIBAYES} \citep{Zwart:2016, Bernardi:2016} to jointly fit the 21-cm spectrum and polynomial foreground. \textsc{HIBAYES} uses a Bayesian framework to explore the signal-plus-foreground posterior probability distribution and evaluate the Bayesian evidence for a given model. Within \textsc{HIBAYES}, we model foregrounds as a 5-term log-polynomial with coefficients $p_i$, and model the 21-cm absorption term as a Gaussian with amplitude $A_{HI}$, central frequency $\nu_{HI}$ and width $\sigma_{HI}$. \ed{We assigned uncertainties based on the radiometer equation, $\sigma = T_{\rm{sys}} / \sqrt{\Delta\nu \tau}$, where $\Delta\nu$=1\,MHz is channel bandwidth, and $\tau$=1\,hr, centered around LST=4\,hr. The system temperature, defined as $T_{\rm{sys}} = T_{\rm{ant}} + T_{\rm{rx}}$, was computed using a receiver temperature $T_{\rm{rx}}$=500\,K.}

\begin{figure*}
\centering
\includegraphics[width=16cm]{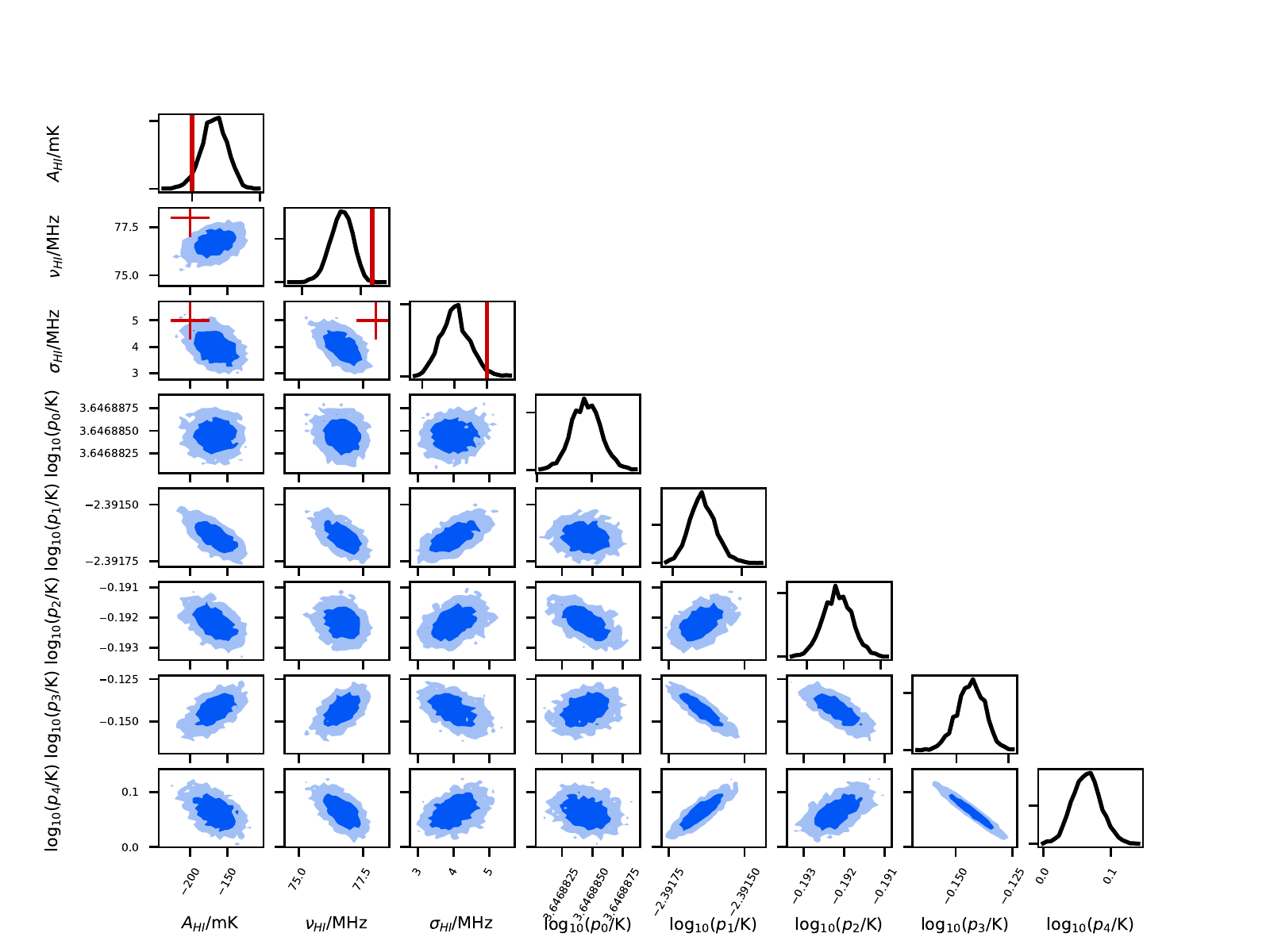}
\caption{Posterior probability distribution, for a N=5 order polynomial and a 200\,mK Gaussian absorption feature at 78\,MHz with 5\,MHz width, fitted to the simulated data using the EDA2 beam pattern with sidelobes suppressed by 20 dB. \ed{True values are indicated in red;} the dark and light shaded regions indicate the 68- and 95-percent confidence regions. }
\label{fig:corner-gauss}
\end{figure*}

The 21-cm feature is not successfully recovered by \textsc{HIBAYES} for the EDA2 beam pattern, or the EDA2 pattern with 10 dB sidelobe suppression. The 21-cm feature is mostly recovered, however, when 20 dB sidelobe suppression is used; \ed{\textsc{HIBAYES} reports {$A_{\rm{HI}}=-169_{-38}^{+75}$\,K, $\nu_{\rm{HI}}=76.8_{-1.9}^{+0.8}$\,MHz, and $\sigma_{\rm{HI}}=4.6_{-1.1}^{+2.2}$ (95\% CI). The global log evidence for the fit is 30.8, which indicates that the fit is `very strong' \citep{Kass:1995}}.} The posterior probability distribution for the EDA2 simulated data is shown in Fig.\,\ref{fig:corner-gauss}. The recovered values for \ed{$\nu_{\rm{HI}}$ are offset from the true values (i.e. the input to simulation), placing them outside the 95\% confidence region. This is likely due to the 1\,MHz channel resolution in our simulation, overfitting due to inflection points caused by the higher-order coefficients of the log-polynomial model, and spectral structure introduced by the sidelobes. In Fig.\,\ref{fig:driftscan-diff}, an local maxima in the residuals is visible close to 78\,MHz; despite the downweighted sidelobe contribution, spectral structure such as this will affect the fit. }

\section{DISCUSSION}

In this article, we investigated whether a beamformed array could be used to measure the 21-cm absorption feature from Cosmic Dawn. Previous studies have suggested that beam chromaticity is a major systematic, and as such a majority of radiometer experiments have favored simple dipole-like antennas. \ed{Our simulations suggest that the $\lambda/D$ evolution of an array beam does not  preclude detection of the Cosmic Dawn. However, sidelobe response is indeed an issue.}

Using the EDA2 as a case study, our simulations suggest that its sidelobes would need to be attenuated by an additional 20 dB in order to detect the 21-cm feature. The mean sidelobe level of the EDA2 is -31\,dB, so a mean sidelobe level of $\sim-50$\,dB is therefore needed. Simulations and theoretical predictions show that the average sidelobe level of a quasi-random array decrease as $1/\sqrt N$, when $N$ is the number of antenna elements. In the SKA-low context, \citet{Sinclair:2015} showed that stations with $O(10^3)$, $O(10^4)$ and $O(10^5)$ elements would have median sidelobe levels of $\sim-32, -42$ and $-52$ dB, respectively; the $-50$\,dB average sidelobe requirement could thus be achieved by an $O(10^5)$ element array. We note that early designs for the SKA included 11,200 antennas \citep{vanArdenne:2012}, which was originally driven by sidelobe requirements for imaging; this was ultimately abandoned in favor of smaller 256-antenna stations so that shorter inter-station baselines were possible. 

A complementary approach to increasing station size for sidelobe suppression is to apply apodization (also known as `windowing') functions when beamforming. These generally have the effect of decreasing sidelobe levels, but also decrease directivity and widen the main beam. Methods such as that of \citet{Taylor:1960} allow for a trade-off between main beam width and sidelobe levels. However, the sidelobe level that a given array can achieve depends upon the number of antennas in the array and how accurate the phasing solutions are---so larger arrays may still be required to reach the $-50$\, dB sidelobe requirement. Efforts with the LWA-SV are already using apodization, but to mitigate the $\lambda/D$ evolution of main beam width \citep{Dilullo:2020, Dilullo:2021}. This has the effect of increasing sidelobe levels; our results suggest that it may be preferable to focus on sidelobe reduction instead, and allow the beam width to change. However, \citet{Dilullo:2020} note that while their approach increases sidelobes, it produces a `smoother' beam by decreasing the depth of nulls in the beam pattern, and does result in improved residuals. While comparison between sidelobe smoothness and null depth is beyond the scope of this work, determining optimal apodization functions for low residuals will be an important area of research for beamformed global signal experiments.

\ed{Here, we simulated a zenith-phased array, in part to avoid effects due to variations in beam patterns at different pointing angles. Our OSKAR simulations do not account for mutual coupling between antenna elements, which causes the beam pattern for each antenna in a closely-packed aperture array to differ. Mutual coupling effects for SKA-low stations have been investigated using Electromagnetic simulation packages \citep{Bolli:2022, Sutinjo:2022}; while considerably more computationally expensive, such simulations offer a more comprehensive method to determine antenna beam patterns. Alternatively, \citet{Kiefner:2021} detail a holographic method to measure an array's antenna beam patterns. These approaches could be used in future work.}

\ed{We also---and purposefully---did not account for the aperture array's beam pattern.} Better results (i.e. lower residuals) are likely achievable if a-priori information about beam pattern and foregrounds are taken into account. For example, the optimal eigenvector methodology of \citet{Hibbard:2020}, which accounts for beam-weighted foregrounds, could be applied to beamformer-based experiments. \ed{The limitations of log-polynomials for foreground modelling have been previously noted, and maximally smooth functions with no inflection points have been proposed as an alternative to polynomial fitting \citep{SathyanarayanaRao:2015, Bevins:2021}.}

An advantage of a beamformed array over a single antenna is that the beam can be steered, and bright astrophysical sources can be used to experimentally determine the beam pattern \citep[as is done in][]{Dilullo:2021}. However, far-out sidelobes are hard to measure, so it may be challenging to reach the required accuracy. Another advantage is that the array itself can be used to measure the sky foreground---particularly if interferometric synthesis imaging is used. We argue that the ability to make {\em{in-situ}} measurements of both the beam and the sky are a promising capability that should be explored further for 21-cm studies.

\ed{A huge amount of care} is taken to characterize and stabilize radiometric 21-cm experiments, as gain fluctuations must be controlled for the radiometer equation to hold over $\sim100$\,hr timescales. A beamformed system can instead use multiple pointings toward astrophysical sources with known frequency spectra for bandpass calibration and to monitor gain variations. Indeed, this approach is used in \citet{Dilullo:2021} at LWA-SV, reaching an RMS of 2.47--5.26\,K using several fitting techniques. While they do not use a drift scan approach, and use a different telescope, this RMS is similar in magnitude to our simulated sidelobe noise.

In summary, we argue that beamforming approaches to 21-cm cosmology are a promising avenue which should be revisited. We encourage further studies and experiments using existing radio arrays with beamforming capability across 45--130\,MHz, and design studies for next-generation aperture arrays with $O(10^4-10^5)$ elements.

\section*{Software}
This work made use of the following software packages: Numpy \citep{Harris:2020}, Astropy \citep{AstropyCollaboration:2013, AstropyCollaboration:2018}, Matplotlib \citep{Hunter:2007}, Healpy \citep{Zonca:2019}, OSKAR \citep{Mort:2017}, PyGDSM \citep{Price:2016}, and HIBAYES \citep{Zwart:2016, Bernardi:2016}.

\begin{acknowledgement}

D. Price thanks R. Wayth for comments on the manuscript, and C. Dillulo for comments and contributions to PyGDSM.

\end{acknowledgement}


\bibliography{references, refs-nonads}

\appendix

\end{document}